\documentclass[aps,prl,twocolumn,showpacs,amssymb]{revtex4}
\usepackage{graphicx}
\usepackage{dcolumn}
\usepackage{bm}
\usepackage{mathptm} 

\renewcommand{\bf}{\rm}

\begin{document}
\date{\today}
\title{How does the chromatin fiber deal with topological constraints?}
\author{Maria Barbi, Julien Mozziconacci and Jean-Marc
Victor}\thanks{Correspondence should be addressed to J.-M. Victor
(email: victor@lptl.jussieu.fr)}
\affiliation{Laboratoire de Physique Theorique des Liquides,
 CNRS UMR 7600}



\begin{abstract}
{In the nuclei of eukaryotic cells, DNA is packaged through several
levels of compaction in an orderly retrievable way that enables the
correct regulation of gene expression. The functional dynamics of this
assembly involves the unwinding of the so-called 30 nm chromatin fiber
and accordingly imposes strong topological constraints. We present a general
method for computing both the twist and the writhe of any winding
pattern.  An explicit derivation is implemented for the chromatin
fiber which provides  the linking number of DNA in eukaryotic
chromosomes. We show that there exists one and only one unwinding
path which satisfies both topological and mechanical constraints that
DNA has to deal with during condensation/decondensation processes.  }
\end{abstract}

\maketitle

In the cells of higher eukaryotes, e.g. animals or plants, meters of
DNA are packaged by means of proteins into a nucleus of a few
micrometer diameter, providing an extreme level of compaction. Coding
sequences ({\em genes}) are therefore dispersed in a mess of folded
DNA and proteins ({\em chromatin}) and should be retrieved at will in
order to enable a correct genetic expression and therefore the cell
survival. This leads to the need of an orderly and dynamically
retrievable structure, which is actually achieved by means of a
chromatin partition into functional domains, each containing one or a
group of genes. In each domain DNA is folded in a hierarchical
structure, including several winding levels. It is first wrapped
around spools of proteins thus forming a ``beads on a string''
assembly, which is in turn folded into a $30$ nm diameter
fiber. This fiber is further organised into a three dimensional
cross-linked network \cite{Poirier}. In this network, two neighboring
nodes are connected by a chromatin fiber loop whose typical length is
about $50\, 000$ base pairs (bp). In order to provide the transcription
machinery with an access to specific genomic regions, the
corresponding loop has to be selectively decondensed, via a reversible
unwinding process that elongates the fiber
\cite{Cirillo}.  The dynamics of this process involves strong
mechanical and topological constraints, the former due to DNA
elasticity \cite{Yao}, the latter due to the conservation of the
linking number $Lk$ of DNA \cite{Crick} in a loop during the
unwinding. Of course, topological constraints could be released by the
intervention of topoisomerases, but it has been shown {\it in vivo}
that chromatin decondensation could take place even when those enzymes
were inhibited \cite{Wright}. Moreover, the classic experiment on {\em
simian virus 40} (SV40) minichromosomes clearly demonstrates that the
linking number is unaffected by the decondensation process {\it in
vitro} \cite{Keller}.

In this Letter we address the issue of how Evolution has dealt with
the extremely difficult problem of finding an efficient winding
pattern fulfilling both mechanical and topological constraints at a
time. To answer this question, we start by giving an analytical
formula for the linking number of DNA in a generic bent fiber. We show
that, despite $Lk$ is known to be a non extensive quantity, it is yet
possible to express it in terms of the mean linking number of the
constitutive elements of the ``beads on a string'' assembly. This
allows us to set about an exhaustive numerical exploration of $Lk$ for
all the possible fiber conformations. By analyzing the results of this
exploration, we are able to infer the existence and the {\em
uniqueness} of a relevant winding/unwinding path satisfying all the
constraints. We show furthermore that this engineering problem is solved at the {\em local} level by the design of the
constitutive elements of the fiber.
\begin{figure}[ht]
\includegraphics[width=.45\textwidth]{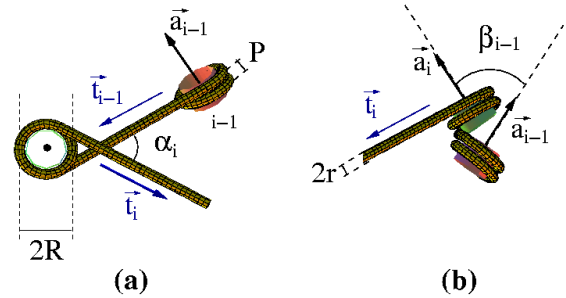}
\caption {Schematic representation of the DNA winding pattern along two
neighbouring nucleosomes in the two-angle model. (a): view down the
NCP axis $\vec{a_{i}}$. (b): view down the linker direction $\vec{t_{i-1}}$.
The angle $\alpha_i$ is the dihedral angle
$(\vec{t_{i-1}},\vec{a_{i}},\vec{t_{i}})$ and $\beta_{i-1}$ the dihedral
angle $(\vec{a_{i-1}},\vec{t_{i-1}},\vec{a_{i}})$ standing for the
twist ({\it modulo} $2\pi$) of the DNA linker. We also indicate
the DNA radius $r\simeq 1.0$ nm, NCP radius $R\simeq 5.3$ nm and NCP pitch
$P\simeq 2.4$ nm.
\label{f1}}
\end{figure}

In a typical chromatin fiber, the beads, called {\em nucleosome core
particles} (NCP), are spaced at intervals of $\sim 200$ bp. Each NCP
contains $\sim 150$ DNA bp, forming $1\frac{3}{4}$ turns of a
left-handed superhelix \cite{Luger}, and two neighboring beads are
connected by $\sim50$ bp stretches of DNA, called {\em linkers}. The unit
of a NCP and a linker is called a {\em nucleosome}. The number of DNA
bp in a nucleosome is known as the {\em repeat length}. In order to
describe the structural polymorphism of a chromatin fiber loop, we use
the original two-angle model (Fig.~\ref{f1}) introduced by Woodcock et
al. \cite{Twoangle}.  In this model, where the linkers are assumed to be straight,
the geometry of a fiber made up of N nucleosomes
is fully characterized by two sets of angles $\alpha_i$ and $\beta_i$ ($i=1\dots N$),
specific of the NCPs and the linkers respectively \cite{Benhaim,Schissel}.
Varying $\alpha_i$ corresponds to changing the number of bp wrapped in the
$i^{\mbox{th}}$ NCP, whereas varying $\beta_i$ corresponds to twisting the
$i^{\mbox{th}}$ linker.

{\it In vivo}, the fiber is most probably homogeneous within a given
loop \cite{Weidemann & Langowski}, i.e. that the distribution of
$\alpha_{i}$ (resp. $\beta_{i}$) is peaked around a mean value $\bar\alpha$ (resp. $\bar\beta$).
During the cell cycle, the biological activity is tuned by
biophysical \cite{Bednar} and biochemical \cite{Allis} parameters,
through a variation of these mean values.
In turn, a variation of $\bar\alpha$ or $\bar\beta$ results in a change of the fiber
internal structure and hence affects its degree of compaction. This
conformational change is necessarily accompanied by a change of the
writhe $W\!r$ and the twist $Tw$ of the DNA double helix.
Consequently, the DNA linking number is {\em a priori} bound to change
according to the White-Fuller theorem $Lk = Tw + W\!r$
\cite{White,Fuller71}. This poses the rather important question of
whether and how the angles $\bar\alpha$ and $\bar\beta$ can be changed, while
maintaining a fixed DNA linking number. To answer to this question, we
set about computing the DNA linking number in a fiber as a
function of the sets $\alpha_i$ and $\beta_i$. This involves the computation of
both the twist and the writhe of the DNA double helix in the fiber.

\begin{figure}[ht]
\includegraphics[width=.25\textwidth]{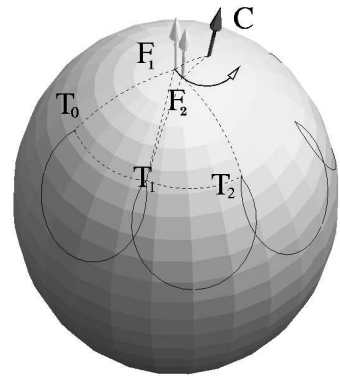}
\caption {Schematic picture of the tangent indicatrix of DNA in a chromatin
fiber with the indication of the particular points used in the
calculations: each solid-line arc corresponds to a nucleosome, with
linker directions given by the connecting points $T_i$.  Points $F_i$
indicate the local fiber axis directions. $C$ is an arbirtarily fixed point.
 All dashed lines indicate geodesic arcs. For
the sake of clarity, scales and lengths have been arbitrarily chosen.
\label{f2}}
\end{figure}

Concerning the computation of the writhe, Fuller's first theorem 
\cite{Fuller78} states that, for a closed curve $\vec
r(s)$ with $s \in [0,L]$, the writhe is related to the {\em signed} area
$A$ enclosed by the tangent indicatrix $T(s)$, namely $W\!r= A/2\pi - 1$
({\em modulo 2}).  $T(s)$ is the curve on the unit sphere traced out
by the tangent vector $\vec t(s)$ to the curve $\vec r(s)$. Fuller's
second theorem \cite{Fuller78} permits us to get rid of the congruence
{\em modulo 2} by computing $W\!r$ from the area $S$ swept out by the
unique shortest geodesic arc from an arbitrarily fixed point $C$ on
the unit sphere to the running point $T(s)$, namely $W\!r = S/2
\pi$. However, the calculation of the writhe of an open curve whose
initial and final tangent vectors $\vec{t}(0)$ and $\vec{t}(L)$ do not
coincide is rather subtle. We follow the procedure recently outlined
by Maggs~\cite{Maggs} to obtain a consistent measure for the writhe by
closing the open tangent indicatrix with a geodesic arc.  This being
rather complicated for a general fiber conformation, it would be
helpful if one could calculate the writhe from the writhes of the
individual nucleosome units. However, the writhe is known to be non
extensive and, as shown by Starostin~\cite{Starostin}, the total
writhe of a curve is given by the sum of the writhes of its parts {\em
plus} the surface of the spherical polygon composed by the set of
geodesics closing each part. In the case of a chromatin fiber, this
non-extensivity may be overcome in the following way. As shown
schematically in Figure~\ref{f2}, the DNA indicatrix $T(s)$ can be
naturally divided into $N$ parts, namely the $N$ nucleosomes, in
correspondence to $\vec{t}_i=\vec{t}(s_i)$, the direction of the
$i^{\mbox{th}}$ linker.  We denote by $T_i$ the point on the unit sphere that
corresponds to $\vec{t}_i$. The total writhe of the curve,
$W\!r$, is then given by the following addition rule:
\begin{equation}
\label{addrule}
W\!r = \sum_{i=1}^{N} W\!r_{i} \,+\, \frac{S_{T_0T_1\dots T_{N}}}{2 \pi} \,.
\end{equation}
$W\!r_{i}$ is the writhe of the $i^{\mbox{th}}$ nucleosome and is
given by the surface $S$
enclosed by the restriction of the tangent indicatrix between
$T_{i-1}$ and $T_{i}$ and the geodesic $T_{i-1}T_i$.
The non-extensive correction $S_{T_0T_1...T_N}$ is the signed area of
the spherical polygon connecting all points $T_i$ that correspond to
the linker directions $\vec{t}_i$. It can be computed as the sum of
the areas of the spherical triangles $T_0T_iT_{i+1}$.  It can also be
expressed with respect to any point $C$ on the unit sphere as $\sum_{i=1}^{N}
S_{CT_{i-1}T_{i}}$ plus a boundary term $S_{CT_{N}T_{0}}$. Up to this
term, the total writhe can be written as $W\!r = \sum_{i=1}^{N}
W\!r_{i}|_{C}$, where
\begin{equation}
W\!r_{i}|_{C} =  W\!r_i + \frac{S_{CT_{i-1}T_{i}}}{2 \pi}
\end{equation}
is the writhe of nucleosome $i$ with respect to point $C$,
corresponding to the writhe of the nucleosome $i$ alone plus the
surface of the spherical triangle $CT_{i-1}T_{i}$ 
.  We see that an appropriate redefinition of the writhe of a
nucleosome permits us to express the total writhe as an extensive
quantity.

\begin{figure*}[ht]

\includegraphics[height=.5\textwidth]{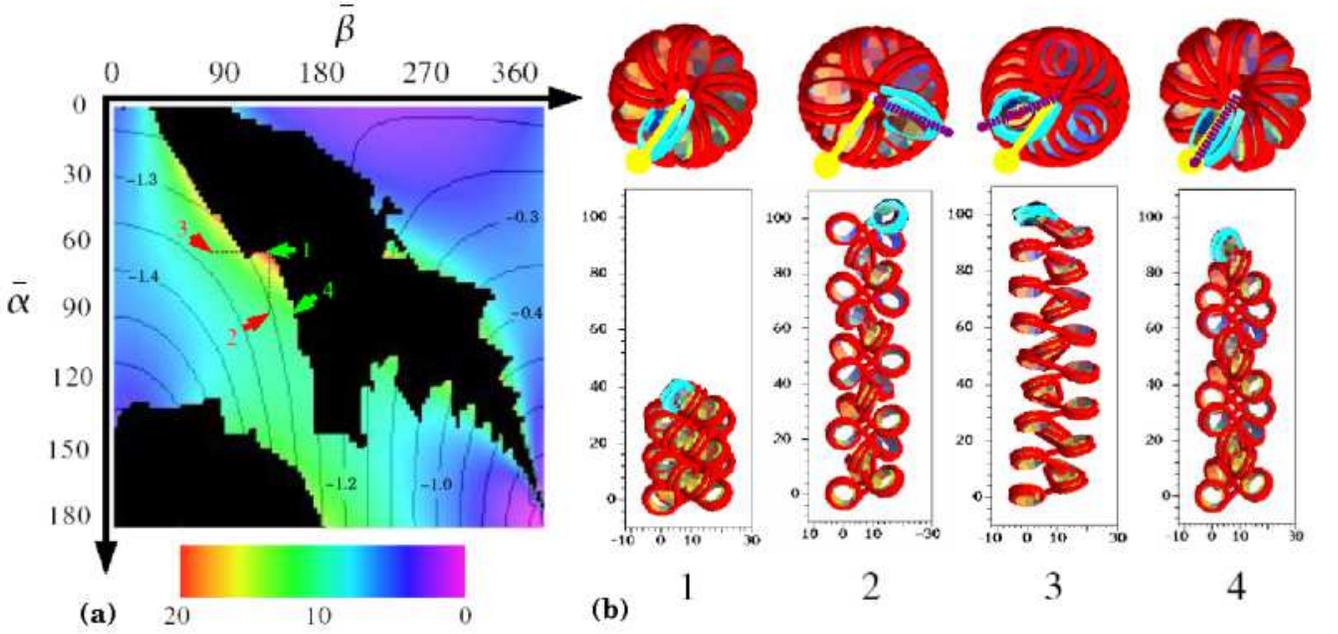}
\caption{(a): solid contour lines (spaced at intervals of $0.05$) 
represent the linking number $Lk(\bar\alpha,\bar\beta)$ computed for $\bar\alpha$
ranging from $0^\circ$ to $180^\circ$, $\bar\beta$ from $0^\circ$ to $360^\circ$ and $\bar
n = 192$ bp.
Color scale gives the fiber compaction defined as the number of
nucleosomes per unit volume $C=1/\pi d R^2$, in units of $NCP/(30$
nm$)^3$.  Black regions correspond to sterically forbidden regions.
Points 1 to 4 are particular reference states: structure 1 is the most
compact fiber structure allowed, paths 1-2, 1-3 and 1-4 correspond
respectively to decondensation at constant $\bar\beta$, constant $\bar\alpha$ and
constant $Lk(\bar\alpha,\bar\beta)$. (b): fiber structures corresponding to points 1 to
4 in (a), top view and frontal view drawn for 24 nucleosomes. For
clarity the first and the last NCP orientations are outlined by a
yellow and a dashed violet line.  The twist change between any two
fiber structures is the angle swept out by the dashed violet line
(namely, $-0.70$ between 1 and 2, $-1.10$ between 1 et 3, $0$ between
1 and 4).
\label{f3}}
\end{figure*}

For the simplest case of a straight regular fiber all the nucleosomes
are evenly dispersed on a regular helix, with the angles $\alpha_i=\alpha$ and
$\beta_i=\beta$ $\forall i$. In this case, though $W\!r_i$ is the same for all
nucleosomes, $W\!r_{i}|_{C}$ varies in general from nucleosome to
nucleosome. Nonetheless there exists a special point $C=F$, defined by
the director of the fiber axis, for which $W\!r_{i}|_{F}$ is
independent of $i$, so that we can define an effective writhe {\em per
nucleosome} $W\!r(\alpha,\beta) = W\!r_{i}|_{F}$, such that $W\!r = N
W\!r(\alpha,\beta)$ \footnote{Note that the same viewpoint $F$ was
introduced by Crick \cite{Crick}.}. We reiterate that the
effective writhe {\em per nucleosome} $W\!r(\alpha,\beta)$ is not the bare
writhe {\em of one individual nucleosome}, the difference being the
surface of the spherical triangle ${FT_0T_{1}}$.

For a generic bent fiber, a local fiber axis $F_i$ can be defined for
each $i$ as the axis of the straight fiber characterized by constant
angles $\alpha=\alpha_i$ and $\beta=\beta_i$ and containing the $i^{\mbox{th}}$ nucleosome.  As
shown in Fig.~\ref{f2}, this allows us to subdivide the surface of the
spherical polygon $T_0T_1...T_N$ into spherical triangles defined with
respect to the points $F_i$, so that the total writhe finally reads
\begin{equation}
\label{decoupage}
W\!r =
 \sum_{i=1}^{N}  W\!r(\alpha_i,\beta_i)  +
\sum_{i=1}^{N-1} \frac{S_{F_{i}T_{i}F_{i+1}}}{2 \pi} + \sum_{i=1}^{N-1}
\frac{S_{CF_{i}F_{i+1}}}{2 \pi}
\end{equation}
with $W\!r(\alpha_i,\beta_i)=W\!r_{i} + \frac{S_{F_{i}T_{i-1}T_{i}}}{2 \pi}$
up to the boundary terms $S_{CF_{N}F_1} + S_{F_{N}T_{N}T_{0}F_1}$.

As concerns the twist of the DNA double helix in the fiber, its
extensivity enables a  more straightforward calculation.  
First of all note that, if the repeat length $n_i$ is allowed to
change from nucleosome to nucleosome, the $i^{\mbox{th}}$ linker will be
relaxed (untwisted) only for a particular equilibrium value
$\beta^{eq}(\alpha_i,n_i)$ of $\beta_i$. Then, let $Tw(\alpha_i,\beta_i)$ be the twist of
the $i^{\mbox{th}}$ nucleosome, with $n_i$ as a repeat length.  At
fixed $\alpha_i$, the variation $\Delta Tw(\alpha_i,\beta_i)$ with respect to $\beta_i$ is
equal to $\Delta \beta_i/2\pi$, whereas, at fixed $\beta_i$, the variation with $\alpha_i$
is equal to the variation of the tortuosity $\tau R \Delta\alpha_i/2\pi$, where $\tau =
2\pi P/(4\pi^2R^2+P^2)$ is the torsion of the double helix axis in the NCP
\footnote{The tortuosity accounts for the twist arising from the
wrapping of the DNA axis into a superhelix, in the absence of
torsional constraint on the double helix itself \cite{Le Bret}.}.  In
order to have an expression for the total twist we just need to know,
therefore, its value on a particular reference state.  
We take as a reference a nucleosome with $146$ bp in the NCP
\cite{Luger} and a relaxed linker. In this state, $\alpha_i=\alpha^0\simeq 60^\circ$, and
 $\beta_i = \beta_i^0= \beta^{eq}(\alpha^0, n_i) = (n_i-146)/h_0$. The twist is then
 $Tw(\alpha^0,\beta_i^0)=146/h + (n_i-146)/h_0$ where $h_0$ (resp. $h$) is the
 helical repeat of the double helix free in solution
 (resp. overtwisted in the NCP) \cite{Wolffe}.  The total twist of the
 DNA double helix in the fiber is therefore
\begin{eqnarray}
\label{twist}
Tw
=\sum_{i=1}^N \tau R \frac{\alpha_i-\alpha^0}{2 \pi}+\frac{\beta_i-\beta_i^0}{2
\pi}+146(\frac{1}{h} - \frac{1}{h_0})+\frac{n_i}{h_0}\,. 
\end{eqnarray}

For the DNA double helix in a straight and relaxed state, the writhe
is zero and the total twist simply amounts to the last term in
Eq.~(\ref{twist}), hence $Lk^{0} = Tw^0 = \sum_{i=1}^N {n_i}/{h_0} = N\, \bar n
/h_0$, where $\bar n$ is the mean repeat length of the fiber.
Therefore, defining the local twist with respect to the relaxed DNA as
$Tw(\alpha_i,\beta_i) = (\tau R ({\alpha_i-\alpha^0})+{\beta_i-\beta_i^0})/{2 \pi}+146({1}/{h} -
{1}/{h_0})$, we obtain from Eq.~(\ref{decoupage}) and
Eq.~(\ref{twist}) the following expression for the DNA {\em excess}
linking number of a generic fiber:
\begin{equation}
\label{LkDNAgeneral}
Lk - Lk^{0}= \sum_{i=1}^{N} Lk(\alpha_{i},\beta_{i})
+ \sum_{i=1}^{N-1} \frac{S_{F_iT_{i}F_{i+1}}}{2 \pi}
+ \sum_{i=1}^{N-1} \frac{S_{CF_iF_{i+1}}}{2 \pi} \,,
\end{equation}
where $Lk(\alpha_{i},\beta_{i})=W\!r(\alpha_i,\beta_i) +Tw(\alpha_i,\beta_i)$ is the local DNA
linking number {\em per nucleosome}.

It is interesting to note that the very last term of
Eq.~(\ref{LkDNAgeneral}), $\sum_{i=1}^{N-1} {S_{CF_iF_{i+1}}}/{2 \pi}$, is
the area $S$ swept out by the geodesic arc from $C$ to the point $F_i$
as $i$ increases. Therefore, this term is nothing but the writhe of
the fiber {\em axis}, $W\!r^{\cal F}\!$.  The excess linking number
$Lk-Lk^0$ is the contribution of the winding pattern of the double
helix to the total linking number $Lk$.  It should therefore be
interpreted as the linking number of the fiber, $L\!k^{\cal F}\!$. The
difference $L\!k^{\cal F}\!-W\!r^{\cal F}\!$ must in turn be
identified with the twist of the fiber around its axis, $Tw^{\cal
F}\!$, in order to satisfy the White-Fuller theorem {\it at the level}
of the fiber $Lk^{\cal F}\!=W\!r^{\cal F}\!+Tw^{\cal F}\!$. This
formal definition of $Tw^{\cal F}\!$ actually matches the intuitive
definition of the fiber twist as the rotation angle of the fiber
``top'' with respect to the fiber ``bottom'', as shown in
Fig.~\ref{f3}(b) where we give the twist variations between different
fibers.

We now evaluate the relative contributions of the three terms
in Eq.~(\ref{LkDNAgeneral}). The first one
can be evaluated in a mean field approximation as $\sum_{i=1}^{N}
Lk(\alpha_{i},\beta_{i})\simeq N \, Lk(\bar\alpha,\bar\beta)$. We have therefore computed
$Lk(\bar\alpha,\bar\beta)$ for $\bar\alpha$ ranging from $0^\circ$ (corresponding to ``open'' NCP
wrapped with 1.5 turns of superhelix \cite{Prunell}) to $180^\circ$
(corresponding to ``closed'' NCP, with 2 complete turns) and $\bar\beta$ from
$0^\circ$ to $360^\circ$ (one period).  A contour plot of our results obtained
for $\bar n=192$ bp is shown in Fig.~\ref{f3}(a).  According to our
definition, $Lk(\bar\alpha,\bar\beta)$ also depends on the mean repeat length
$\bar n$ trough the term $\bar\beta^0/2\pi=(\bar n-146)/h_0$ ({\em modulo
$2\pi$}).
Anyway, changing $\bar n$ simply shifts the levels of the contour
lines without affecting their shapes. The last two terms of
Eq.~(\ref{LkDNAgeneral}) arise from the bending of the fiber. Taking
into account that the mean radius of curvature of the fiber axis is
equal to the fiber persistence length, which is $\simeq 30$ nm
\cite{Benhaim}, it is possible to show that their net contribution
{\em per nucleosome} never exceeds $0.01$.
We can therefore conclude that a $Lk$--conserving path in the
$(\alpha,\beta)$ plane practically never deviates from a given contour line.

We finally come to the issue of whether  a biologically
relevant $Lk$--conserving decondensation path in the $(\bar\alpha, \bar\beta)$ plane
can actually exist. We know that such a path must remain close
to a contour line.  Moreover, it should be capable of transforming a
highly compact structure into a decondensed one. A fiber
compaction map is displayed in color scale in Fig.~\ref{f3}(a).  An
exaustive scan of this map shows that there is only one small region,
around point 1, which provides a degree of compaction high enough to
match the maximal densities observed {\it in vitro},
i.e. $\sim6\,NCP/10$ nm (red in the color scale). Hence, there can be
only one $Lk$--conserving decondensation path, namely the contour line
going through point 1 and leading to point 4.  It is clear from
Fig.~\ref{f3}(a) that varying only $\bar\alpha$ (leading to point 2) or only
$\bar\beta$ (leading to point 3) is forbidden at constant $Lk$.

Moreover, energy considerations have to be taken into account.  As
shown by Langowski in a recent simulation \cite{Langowski}, the
internucleosomal interactions play a minor role in the control of the
fiber compaction while the elastic energy (bending and twisting) of
the linkers plays the major part. It comes out that the linkers should
remain relaxed all over the decondensation path. First, in order to
have relaxed linkers at point 1, $\bar\beta$ must be equal to
$\beta^{eq}(\alpha^0,\bar n)$: this leads to select a special value for the
mean repeat length $\bar n$, equal to 192 bp (modulo $10.6$ bp).
Secondly, during the decondensation (path 1--4), $\bar\alpha$ varies from
$\sim 60^\circ$ to $\sim 90^\circ$. This variation corresponds to the wrapping of $\sim
10$ bp of the linker onto the protein spool. As the DNA helical repeat
is lower in the NCP ($h=10.2$ bp/turn) than in the linkers ($h_0=10.6$
bp/turn), this wrapping is accompanied by a change of the equilibrium
value $\beta^{eq}(\bar\alpha,\bar n)$ from $120^\circ$ to $140^\circ$. Remarkably,
along the decondensation path 1--4, $\bar\beta$ varies as well from
$120^\circ$ to $140^\circ$.  Hence $\bar\beta=\beta^{eq}(\bar\alpha,\bar n)$ all over the
path 1--4, provided that $\bar n =192 bp$, which is the repeat length measured in HeLa cells.

Our results provide a new way
of understanding the quantization of the
repeat lengths \cite{Widom}. The average repeat length has to be
selected, e.g. thanks to nucleosome positioning sequences, in order to
fix the mean nucleosome orientation $\beta^{eq}(\bar \alpha, \bar n)$. This
particular orientation not only provides the maximal fiber compaction
with relaxed linkers \cite{Widom}, but also enables the global
conservation of the linking number of the fiber during its
winding/unwinding process driven by the nucleosome internal dynamics itself.

\acknowledgements{
We thank A. Lesne, A. Paldi, A. Prunell and C. Lavelle for fruitful
discussions, R. Chitra for a very careful reading of the paper, and
M. Quaggetto for technical support.}

\bibliographystyle{unsrt}

\end{document}